\documentclass[aps,preprint,prb]{revtex4}
\usepackage{graphicx}
\usepackage{epsfig}
\usepackage{amsmath}
\usepackage{graphics}
\begin{document}

\author{M. F. Gelin}
\author{D. S. Kosov}

\affiliation{Department of Chemistry and Biochemistry,
 University of Maryland, 
 College Park, 
 20742, 
 USA}

\title{Molecular reorientation in hydrogen-bonding liquids:
through algebraic $\sim t^{-3/2}$ relaxation toward exponential decay}

\begin{abstract}
We present a model for the description of orientational relaxation
in hydrogen-bonding liquids. The model contains two relaxation parameters
which regulate the intensity and efficiency of dissipation, as well
as the memory function which is responsible for the short-time relaxation
effects. It is shown that the librational portion of the orientational
relaxation is described by an algebraic $\sim t^{-3/2}$ contribution,
on top of which more rapid and non-monotonous decays caused
by the memory effects are superimposed. The long-time behavior of the orientational
relaxation is exponential, although non-diffusional. It is governed
by the rotational energy relaxation. We apply the  model to interpret
recent molecular dynamic simulations and polarization pump-probe experiments  
on $HOD$ in liquid $D_{2}O$  [C. J. Fecko et al, J. Chem.
Phys. 122, 054506 (2005)].
\end{abstract}
\maketitle

\section{Introduction}

The recent advances in nonlinear ultrafast polarization-sensitive
spectroscopy \cite{zew01} make it possible to monitor molecular
rotation in hydrogen-bonding liquids  in real
time \cite{tok04,tok05,fay05,bak05,bak00,fay04,lau98,kei99,wie04,bra02}.
Due to the enormous complexity of the problem, which is exacerbated by many-body effects and multitudes  of the time scales involved, experimental data alone are insufficient 
for understanding the underlying  dynamics. 
Nowadays, molecular dynamic has become a standard tool for studying molecular
reorientation in liquids \cite{AlTi}. Furthermore, one can even use \textit{ab
initio} molecular dynamics (in which the density functional
theory is invoked to describe molecular electronic structure and inter- and intramolecular
forces are calculated on-the-fly) \cite{marx03,par96} or centroid
molecular dynamics (which accounts for quantum effects/corrections)
\cite{kus04}. On the other hand, there exists a  plenty of "old" phenomenological models of molecular reorientation in gases and 
liquids. We mention the small-angle rotational diffusion model \cite{fav60,hub70,hun70},
the jump diffusion model \cite{val73,cuk72,cuk74,gel98}, the friction model
\cite{ste63,McCo,LB84}, the Gaussian cage model \cite{ste84,ste85},
the itinerant oscillator model \cite{gri84}, along with the more sophisticated
memory function approach (\cite{BurTe,gri84,ste84a,kiv88,det80,key72}
and references therein), the extended diffusion models \cite{gor66,sack,rid69,ste72,bra78,McCl77,bul84,gel98a,con76},
the rotational Fokker-Planck equation \cite{sack,rid69,hub72,McCo,mor82,McCl87,gel96a},
the confined rotator model \cite{kal81}, the Steele model \cite{ste81},
the Keilson-Storer model (KSM) \cite{sack,BurTe,gel97,gel97a,gel00,gel96,gel01},
the fluctuating/stochastic cage model \cite{fre95,BurTe,bur83,pol04,pol05},
and the generalized Langevin equations/normal mode approach \cite{str00,BurTe,gri84,ste84a}. Furthermore, the model has been elaborated
\cite{bra02}, which accounts for the effects of rotation-vibration
coupling in ensembles of hydrogen-bonding molecules on the time-resolved
pump-probe signals. Very recently, the generalized jump model of water reorientation has been suggested \cite{hyn06}. 
The models, of course, rely upon a simplified picture of molecular rotation. However, in contrast with molecular dynamics simulations, they  
get a deep insight into physics of molecular reorientation, give a clear perception of rotational relaxation and provide us with explicit formulas for the pertinent correlation functions (CFs).

The aim of the present paper is to develop a simple and physically sound model of molecular reorientation in hydrogen-bonding liquids. The model is intended  to supply experimentalists with a simple theory to  interpret and to fit  their data and to clarify
the interconnection of orientational relaxation and hydrogen bond making/breaking processes. In  hydrogen-bonding liquids, the angular momentum
CFs exhibit pronounced oscillations and orientational CFs (OCFs) display a rapid short-time  decay followed by a slower (sometimes oscillatory)
pattern which transforms gradually into a monotonous exponential relaxation. By incorporating the proper description of the memory effects into the
KSM framework, we developed a non-Markovian generalization of thereof, NKSM. Within the NKSM, we  derived analytical expressions
for the angular momentum and energy CFs, as well as simple recursive expressions for OCFs.

The paper is structured as follows. The NKSM is formulated in Sec.II. 
The explicit expressions for the angular momentum CF, rotational energy CF and OCFs are presented and discussed in Sec.III. 
Sec.IV contains illustrative calculations of various NKSM CFs and comparisons
with the results of molecular dynamic simulations and polarization pump-probe experiments  
on $HOD$ in $D_{2}O$ at a room temperature \cite{tok05}. A brief summary of the main findings can
be found in Sec. V. Appendix A contains the explicit formulas for
the calculation of spherical and linear rotor OCFs within the NKSM.
Analytical expressions for OCFs in a particular case of ``perfect''
librations are obtained and discussed in Appendix B. 

A few words about the notation and conventions. (i) The reduced variables
are used throughout the article: time, angular momentum and energy
are measured in units of $\sqrt{I/(k_{B}T})$, $\sqrt{Ik_{B}T}$ and
$k_{B}T$, respectively. Here $k_{B}$ is the Boltzmann constant,
$T$ is the temperature, and $I$ is a characteristic moment of inertia
of the molecule so that $\tau_{r}=\sqrt{I/(k_{B}T})$ is the averaged
period of free rotation. (ii) All  Laplace-transformed operators
are denoted by tilde, viz. $\tilde{f}(s)=\int_{0}^{\infty}dt\exp\{-st\} f(t)$
for $\forall$ $f(t)$. (iii) Repeated dummy Greek indexes imply summation
over $x,\, y$ and $z$.

\section{The model}

We start with a formally exact Zwanzig-type master equation, which
can be derived from the general $N$-particle rotation-translational
Liouville equation by applying the projection operator technique \cite{fre75,eva78,gel98}

\begin{equation}
\partial_{t}\rho(\mathbf{J},\mathbf{\Omega},t)=-i\hat{\Lambda}(\mathbf{J},\mathbf{\Omega})\rho(\mathbf{J},\mathbf{\Omega},t)-\int_{0}^{t}dt'\hat{C}(\mathbf{J},\mathbf{\Omega},t-t')\rho(\mathbf{J},\mathbf{\Omega},t').\label{kin1}\end{equation}
 Here $\rho(\mathbf{J},\mathbf{\Omega},t)$ is the single particle
probability density function, $\mathbf{J}$ is the angular momentum
in the molecular frame, $\mathbf{\Omega}$ are the Euler angles which
specify orientation of the molecular frame with respect to the laboratory
one. The free-rotor Liouville operator consists of the two contributions,
\begin{equation}
\hat{\Lambda}(\mathbf{J},\mathbf{\Omega})=\hat{\Lambda}_{\mathbf{\Omega}}+\hat{\Lambda}_{\mathbf{J}},\label{str}\end{equation}
which describe, respectively, the angular momentum driven reorientation
and the angular momentum change during free rotation:\begin{equation}
\hat{\Lambda}_{\mathbf{\Omega}}=I_{\alpha}^{-1}J_{\alpha}\hat{L}_{\alpha},\,\,\,\hat{\Lambda}_{\mathbf{J}}=-i\varepsilon_{\alpha\beta\gamma}I_{\beta}^{-1}J_{\alpha}J_{\beta}\partial_{J_{\gamma}}.\label{str1}\end{equation}
 $I_{\alpha}$ are the main moments of inertia, $\hat{L}_{\alpha}$
are the angular momentum operators in the molecular frame. For linear
and spherical rotors, $\hat{\Lambda}_{\mathbf{J}}\equiv0$. 

The relaxation operator $\hat{C}$ assumes the form\begin{equation}
\hat{C}(\mathbf{J},\mathbf{\Omega},t)=\hat{C}(\mathbf{J})g(t),\label{LiHo}\end{equation}
$g(t)$ being the memory function which is normalized to unity, $\int_{0}^{\infty}dtg(t)=\tilde{g}(0)=1$.
All the formulas derived in the present paper are valid for any functional
form of $g(t)$. Since a simple exponential memory function is known
to exaggerate oscillatory effects in the angular momentum CF and OCFs
(see, e.g., \cite{det80,ste84a,gel97a,LB84}), the two-exponential
memory function will be adopted for making all specific calculations,
viz.,

\begin{equation}
g(t)=\sigma\lambda_{1}\exp\{-\lambda_{1}t\}+(1-\sigma)\lambda_{2}\exp\{-\lambda_{2}t\},\label{g(t)}\end{equation}
\begin{equation}
\tilde{g}(s)=\frac{\sigma\lambda_{1}}{s+\lambda_{1}}+\frac{(1-\sigma)\lambda_{2}}{s+\lambda_{2}}.\label{g(s)}\end{equation}
The parameters $\lambda_{i}$ regulate the memory effects of the two
contributions, and $\sigma$ controls their relative significance.
If we let both $\lambda_{1}$ and $\lambda_{2}$ tend to infinity, then $g(t)\rightarrow\delta(t)$
and the Markovian limit is recovered. 

Since molecules are massive inertial particles, the relaxation operator
$\hat{C}(\mathbf{J})$ is assumed to be $\mathbf{\Omega}$-independent
\cite{foot3}. This is tantamount to the statement that molecular reorientation
is driven by the (time-dependent) angular momenta, whose relaxation,
in turn, is governed by operator (\ref{LiHo}). This assumption is
consistent with classical molecular dynamics simulations,
in which one integrates equations of motion of the kind \[
\partial_{t}D^{j}(\Omega)=-i\hat{\Lambda}_{\mathbf{\Omega}}D^{j}(\Omega),\,\,\,\partial_{t}\mathbf{J}=-i\hat{\Lambda}_{\mathbf{J}}\mathbf{J}+\mathbf{N},\]
$D^{j}(\Omega)$ being the Wigner D-functions \cite{var89} and $\mathbf{N}$
being the torque acting on a chosen molecule from its neighbours.
As is demonstrated below, $\mathbf{N}$ is essentially non-Gaussian
and non-Markovian. 

The operator $\hat{C}$ can further be represented in the general
form \cite{BurTe}\begin{equation}
\hat{C}(\mathbf{J})\rho(\mathbf{J},\mathbf{\Omega},t)=-\nu\{\rho(\mathbf{J},\mathbf{\Omega},t)-\int d\mathbf{J}'T(\mathbf{J}|\mathbf{J}')\rho(\mathbf{J}',\mathbf{\Omega},t)\}.\label{Ci}\end{equation}
 The rate $\nu$ determines the dissipation strength, and the relaxation
kernel $T$ obeys the normalization 

\begin{equation}
\int d\mathbf{J}T(\mathbf{J}|\mathbf{J}')=1\label{norm}\end{equation}
and the detailed balance \begin{equation}
T(\mathbf{J}|\mathbf{J}')\rho_{B}(\mathbf{J}')=T(\mathbf{J}'|\mathbf{J})\rho_{B}(\mathbf{J}),\label{DetBal}\end{equation}

\begin{equation}
\rho_{B}(\mathbf{J})=(2\pi)^{-3/2}\exp\{-J_{\alpha}^{2}/(2I_{\alpha})\}\label{Boltz}\end{equation}
being the equilibrium rotational Boltzmann distribution. 

To proceed further, we adopt the KSM parametrization of the relaxation
kernel \cite{BurTe,KS}:

\begin{equation}
T(\mathbf{J}|\mathbf{J}')=\prod_{a=x,y,z}T_{a}(J_{a}|J_{a}'),\label{KS}\end{equation}
\[
T_{a}(J_{a}|J_{a}')=[2\pi I_{a}(1-\gamma_{a}^{2})]^{-1/2}\exp\{-(J_{a}-\gamma_{a}J_{a}')^{2}/[2I_{a}(1-\gamma_{a}^{2})]\}.\]
 Here the parameters $-1\leq\gamma_{a}\leq1$ determine the relaxation
mechanisms. When $\gamma_{a}=1$, then $T_{a}(J_{a}|J_{a}')$$=\delta(J_{a}-J_{a}')$
and $\hat{C}(\mathbf{J})=0$. The Fokker-Planck relaxation operator
is recovered in the limit $\gamma_{a}\rightarrow1$, $\nu\rightarrow\infty$,
$\nu(1-\gamma_{a})\rightarrow\nu_{a}=\mathrm{const}$. If $\gamma_{a}=0,$
intermolecular interactions are so strong that they {}``immediately''
restore an equilibrium Boltzmann distribution in the molecular ensemble
($T(\mathbf{J}|\mathbf{J}')\rightarrow\rho_{B}(\mathbf{J})$). Therefore, the
KSM contains the J-diffusion model \cite{gor66,sack,rid69,ste72,bra78,McCl77,bul84,gel98a,con76}
and the rotational Fokker-Planck equation \cite{sack,rid69,hub72,McCo,mor82,McCl87}
as  special cases. 

By letting $\gamma_{a}=-1$, one gets $T_{a}(J_{a}|J_{a}')$ $=\delta(J_{a}+J_{a}')$.
Thus the magnitude of the angular momentum is preserved but its direction
is reversed. This regime ($\gamma_{a}\simeq-1$, a molecule rotates
back and forth within the cage formed by its nearest neighbors) is
expected to be of particularly relevance for hydrogen-bonding liquids.
Such physical picture of molecular reorientation in liquids is inherent
in a number of theoretical approaches, in which the influence of the
nearest neighbors on the selected molecule is modeled by external
potentials with several minima 
\cite{ste84,ste85,kal81,kus77,kam81,deb88,eva95,bur94,lit72,fre55},
or by fluctuating torques and structures \cite{gri84,AlTi,fre95}. Within
the present approach, the fluctuating cage potential is not introduced
explicitly, but its influence is taken into account dynamically  via kernel (\ref{KS}). 

Before embarking at particular calculations, it is useful to estimate
values of the parameters $\nu,\,\gamma_{a}$ (Eq. (\ref{KS})) and
$\lambda_{i}$ (Eq. (\ref{g(t)}) for hydrogen-bonding liquids. Since
the molecules are assumed to undergo hindered librations, one expects
$\nu\gg1$ and $\gamma_{a}\sim-1$. The memory effects are supposed
to be quite significant ($\lambda_{i}\sim1$).

\section{Correlation functions}

After the explicit form of the relaxation operator (\ref{KS}) has
been determined, the master equation (\ref{kin1}) can be invoked
to calculate any rotational and/or orientational CF of interest. 
To study  the evolution of any quantity, which 
depends solely on the angular momentum, we integrate Eq. (\ref{kin1})
over $\mathbf{\Omega}$ and obtain the reduced master equation

\begin{equation}
\partial_{t}\rho(\mathbf{J},t)=-i\hat{\Lambda}_{\mathbf{J}}\rho(\mathbf{J},t)-\int_{0}^{t}dt'g(t-t')\hat{C}(\mathbf{J})\rho(\mathbf{J},t').\label{kinJ}\end{equation}
OCF of the rank $j$ is defined through the Wigner D-functions as
follows:

\begin{equation}
G^{j}(t)\equiv<D^{j}(\Omega(t))D^{j}(\Omega(0))^{-1}>\equiv\int d\mathbf{J}G^{j}(\mathbf{J},t).\label{OCF}\end{equation}
 After the insertion of the above definition into Eq. (\ref{kin1})
one obtains the following equation: \begin{equation}
\partial_{t}G^{j}(\mathbf{J},t)=-i(\Lambda_{\mathbf{\Omega}}+\hat{\Lambda}_{\mathbf{J}})G^{j}(\mathbf{J},t)-\int_{0}^{t}dt'g(t-t')\hat{C}(\mathbf{J})G^{j}(\mathbf{J},t').\label{OCF1}\end{equation}
Here operator $\Lambda(\mathbf{J})$ is determined by Eqs. (\ref{str})
and (\ref{str1}), in which the angular momentum operators $\hat{L}_{\alpha}$
are replaced by their matrix elements $L_{\alpha}^{j}$ over the D-functions:

\begin{equation}
(L_{x}^{j})_{kl}\pm i(L_{y}^{j})_{kl}=\delta_{k,l\mp1}\{(j\pm l)(j\mp l+1)\}^{1/2},\,\,(L_{z}^{j})_{kl}=l\delta_{kl};\,\,-j\leq k,l\leq j.\label{Jxyz}\end{equation}
Eqs. (\ref{kinJ}) and (\ref{OCF1}) can be solved numerically in
case if a general asymmetric top molecule. To make the presentation
simpler, we restrict ourselves to the consideration of spherically
symmetric molecules. The corresponding formulas are much more elucidating
and convenient to analyse, since the relaxation operator $\hat{C}(\mathbf{J})$
is described by only two dynamic parameters (the intensity, $\nu$,
and the efficiency, $\gamma$, of dissipation) and a spherical molecule
possesses a single moment of inertia, $I$. This theory can be applied
to asymmetric tops as well. Indeed, in the hindered rotation limit
($\tau_{J}\ll1$), a molecule librates back and forth in its cage,
and every single libration reorients the molecule to a small angle.
Using the explicit form of the operator $\hat{\Lambda}_{\mathbf{\Omega}}$
(\ref{str1}), it is easy to demonstrate that the averaged inertial
reorientation angle of an asymmetric top around its $z$-axis equals
$(t/\tau_{r})^{2}\ll1$, where \begin{equation}
\tau_{r}=\sqrt{I/(k_{B}T}),\,\,\, I^{-1}=(I_{x}^{-1}+I_{y}^{-1})/2.\label{Ixy}\end{equation}
 This corresponds to the rotation of the spherical molecule with the
effective moment of inertia $I$. Thus, all the formulas obtained
below can be used for asymmetric top molecules by making use of the
substitution (\ref{Ixy}), provided one is interested in the orientational
relaxation of the tensor with nonzero components along the molecular
$z$-axis. If the quantity under study possesses nonzero components
along several axes of the main moments of inertia (like $OH$ stretch
in $HOD$), the above analysis remains true if the effective moment
of inertia $I$ is modified accordingly. On the contrary, free inertial
rotation of spherical, symmetric and asymmetric tops is very different
\cite{ste69,blo86,lei86}.

Starting from Eq. (\ref{kinJ}), it is straightforward to derive the explicit formulas for the Laplace
transformations of the angular momentum and rotational energy CFs:

\begin{equation}
C_{J}(t)=\frac{\left\langle \textbf{JJ}(t)\right\rangle }{\left\langle \textbf{J}^{2}\right\rangle },\,\,\,\tilde{C}_{J}(s)=\frac{1}{s+\nu_{J}\tilde{g}(s)},\label{CJ}\end{equation}
\begin{equation}
C_{E}(t)=\frac{\left\langle \textbf{J}^{2}\textbf{J}^{2}(t)\right\rangle -\left\langle \textbf{J}^{2}\right\rangle ^{2}}{\left\langle \textbf{J}^{4}\right\rangle -\left\langle \textbf{J}^{2}\right\rangle ^{2}},\,\,\,\tilde{C}_{E}(s)=\frac{1}{s+\nu_{E}\tilde{g}(s)}.\label{CE}\end{equation}
 Here the rates \begin{equation}
\nu_{J}=\nu(1-\gamma),\,\,\nu_{E}=\nu(1-\gamma^{2})\label{NuJE}\end{equation}
 determine the angular momentum and rotational energy integral relaxation
times:\begin{equation}
\tau_{J}=\int_{0}^{\infty}dtC_{J}(t)=\tilde{C}_{J}(0)=\nu_{J}^{-1},\label{TauJ}\end{equation}
\begin{equation}
\tau_{E}=\int_{0}^{\infty}dtC_{E}(t)=\tilde{C}_{E}(0)=\nu_{E}^{-1}.\label{TauE}\end{equation}

As to the OCFs, the solution of Eq. (\ref{OCF1})
can be given in terms of recursive relationships (or, equivalently,
in terms of continued fractions) for any value of the relaxation parameters $\nu$ and $\gamma$, as
well as for any memory function $g(t)$, see Appendix A. 

\section{ Illustrative calculations}
In this Section, we present and discuss the results of representative calculations of the angular momentum CFs (\ref{CJ}), energy CFs (\ref{CE}), and OCFs  (\ref{Gb0}) - (\ref{sisi}) within the NKSM.      
The free rotation period (\ref{Ixy}) has been taken as $\tau_{r}=\sqrt{I/(k_{B}T})=93$fs.
This corresponds to $HOD$ at  $T=300$ K ($I_{x}=2.63$, $I_{y}=1.85$,
$I_{z}=0.72$ $a.m.u.\times \AA^{2}$). 

We start from the angular momentum (\ref{CJ}) and energy (\ref{CE})
CFs. If the two-exponential memory function (\ref{g(t)}) is used, the
Laplace images $\tilde{C}_{J}(s)$ and $\tilde{C}_{E}(s)$ can be inverted into the time domain by solving the pertinent cubic equation. The CFs in the time domain are thus determined by
linear combinations of one real and two complex conjugated exponentials.
The results of representative calculations are depicted in Fig.
1. A decrease of the second memory parameter, $\lambda_{2}$, causes
a typical transformation of the angular momentum CF, which reflects
a passage from simple to hydrogen-bonding liquids \cite{LB84}. Since
CFs $C_{J}(t)$ attain negative values, their actual decay occur at
a timescale of several hundreds of femtosecond, which is much longer
than the integral relaxation time $\tau_{J}=\nu_{J}^{-1}=4.65$ fs.
The rotational energy CFs are monotonous and decay at a much longer
time scale of $10\div15$ ps, which is a direct manifestation of the
librational motion ($\gamma\sim-1$, $\tau_{J}\ll\tau_{E}$). 

Let us turn to the study of orientational relaxation. 
To get a qualitative feeling
of the influence of the relaxation efficiency $\gamma$ on molecular
reorientation, let us concentrate on the long-time behavior of OCFs
and consider the Markovian limit, $g(t)\rightarrow\delta(t)$. As
has been established in \cite{gel96,gel01}, the KSM predicts that
the smaller is $\gamma$, the slower is the OCF decay. Thus, for a
fixed angular momentum relaxation rate $\nu_{J}$, the rotational
Fokker-Planck equation ($\gamma=1$) predicts the most rapid orientational
relaxation, while the limit of perfect librations ($\gamma=-1$) corresponds
to the slowest orientational relaxation. This is clearly seen in Fig.
2. 

As is demonstrated in Appendix B, Eq. (\ref{OCF1}) can be solved
analytically for OCFs in case of perfect forward-backward librations ($\gamma=-1$
) in the hindered rotation limit ($\nu\gg1$): 

\begin{equation}
G^{j}(t)=\frac{1}{2j+1}\left(1+2\sum_{k=1}^{j}\left[1+\frac{k^{2}}{\nu}t\right]^{-3/2}\right).\label{Gt2}\end{equation}
 On the scale of Fig. 2, the exact solution of Eq. (\ref{OCF1}) in
the limit of $\gamma=-1$ and the approximate one, which is delivered
by Eq. (\ref{Gt2}), are indistinguishable. Thus the in-cage librations
manifest themselves through a slow algebraic $t^{-3/2}$ decay of
the OCF. This behavior is caused by the angular momentum reversion
($\gamma=-1$) and has nothing in common with long-time hydrodynamic
tails of the angular velocity CFs of Brownian particles (see \cite{fel97,mas97,AlTi}
and references therein) \cite{foot2}. It is interesting to point
out that the OCFs calculated within the M- \cite{bha85,McCl77} and
E- \cite{gel98a} diffusion models exhibit similar ($\sim t^{-3/2}$
) long time tails, which are commonly regarded as unphysical. Nonetheless,
these models were successfully invoked to reproduce {}``experimental''
OCFs, which were obtained through the inversion of IR and Raman spectra
into the time domain \cite{bha85,dre79,gel98a,jon81,rot75}. The present
consideration reveals that this success is not accidental, since the
conservation of the magnitude of the angular momentum (M- diffusion)
or rotational energy (E- diffusion) mimics the description of in-cage
librations via the KSM kernel (\ref{KS}), which in the limit $\gamma\sim-1$
conserves both of these quantities. 

Fig. 3 illustrates the influence of the memory effects, which modify
the short-time behavior of the OCFs. The mechanism of this influence
is uncovered by the general expression \cite{McCo}\begin{equation}
G^{j}(t)\approx1-j(j+1)\int_{0}^{t}dt'(t-t')C_{J}(t'),\label{GtShort}\end{equation}
which relates the short-time OCF behavior with the time evolution
of the angular momentum CF. That is why the parameters which have
been used for the calculation of the OCFs in Fig. 3 and the angular
momentum CFs in Fig. 1a are the same. Generally, the memory effects
speed up the short-time OCF decay. If the angular momentum CF is highly
oscillatory, these oscillations show up in OCFs also (compare dotted
lines in Figs. 1a and 3). 

OCFs depicted in Fig. 3 look qualitatively similar to those obtained
via computer simulations \cite{AlTi,ski03,tok04,tok05,kus04,kus94}.
A more quantitative comparison of the simulated and NKSM OCFs is presented
in Fig. 4. It depicts the results of molecular dynamic simulations 
of the first (upper dotted line) and second (lower dotted line) rank OCFs
for $HOD$ in $D_{2}O$ at a room temperature using the SPC/E model for water\cite{tok05} along with the fits obtained within the NKSM (full lines). Detail of the molecular dynamics simulation protocol can be found in ref.\cite{tok04,eav04}

The present theory is seen to reproduce the simulated OCFs quite well.
We were unable, however, to fit the first and second rank OCFs by
the same set of the NKSM parameters. This is not unexpected: OCFs
of different ranks are affected by the cage potentials in a different
way, which is determined by the potential symmetry \cite{fre95}.
The dielectric friction, which governs orientational relaxation in
polar systems is also known to be rank-dependent \cite{bag94}. The
values of $\nu$, $\gamma$, $\lambda_{i}$ and $\sigma$ which are
extracted from the angular momentum CF deliver, in fact, certain averaged
values. It is nonetheless quite remarkable that the parameters which
have been used for the calculation of the second rank OCF are identical
to those which have been used for the computation of the angular momentum
CF (Fig. 1a, full line). This latter CF looks very similar to the
simulated one \cite{kus04}. 

Fig. 5 shows the normalized anisotropy extracted from the pump-probe
signal \cite{tok05} (dotted line). It deviates quite significantly
from the simulated second-rank OCF (our best fit to this OCF is plotted
here for the sake of comparison). As has been pointed out in \cite{tok05,tok04,ski03},
simulations normally predict faster, in comparison with experiment,
anisotropy decays. This is caused, perhaps, by ignoring either the
water polarizability \cite{ber02,ich99} or quantum effects \cite{kus04}.
We are not attempting to resolve this controversy here. Note merely
that the experimental anisotropy can be fitted quite well within the
NKSM (full line), with the set of parameters which predict more {}``librational''
reorientation ($\gamma$ is closer to $-1$) and more pronounced memory
effects ($\lambda_{i}$ is smaller). 

Let us return back to Fig. 4. The OCFs calculated within the standard
diffusion equation, \begin{equation}
G^{j}(t)=\exp\{-j(j+1)\tau_{J}t\},\label{dif}\end{equation}
 which reproduces the long-time limit of the first cumulant formula,
\begin{equation}
G^{j}(t)=\exp\{-j(j+1)\int_{0}^{t}dt'(t-t')C_{J}(t')\},\label{GtShort1}\end{equation}
are seen to deviate significantly from the simulated/NKSM OCFs. Despite
the simulated/NKSM OCFs do exhibit the long-time exponential behavior,
and despite the rotational motion is definitely hindered ($\tau_{J}\ll1$),
the diffusion equation cannot reproduce these OCFs. The failure of
the diffusion equation has quite an evident explanation. If we assume
that the molecules undergo {}``perfect'' librations ($\gamma=-1$
within the present theory), then the long-time behavior of the OCF
in the hindered rotation limit is described by Eq. (\ref{Gt2}) rather
than by the small-angle diffusion equation (\ref{dif}). Since the
actual librations are not {}``perfect'' ($\gamma$ is close to but
less than $-1$) one expects deviations from Eq. (\ref{Gt2}). As is
seen from Fig. 6, this is indeed the case. This Figure reproduces
the Markovian limits of the KSM OCFs from Figs. 4 and 5, along with
their counterparts calculated via Eq. (\ref{Gt2}). The short-to-intermediate-time
resemblance of the OCFs is quite remarkable. 

On the other hand, the long-time behavior of the simulated/KSM OCFs
is seen to be exponential and is not reproduced by Eq. (\ref{Gt2}).
This hints at a possibility that rotational energy relaxation might
be responsible for the long time exponential decay of the simulated/NKSM
OCFs. This hypothesis is corroborated by the observation that the
fit of the simulated and NKSM OCFs of the first and second rank (Fig.
4) delivers different values of the relaxation parameters $\nu$ and
$\gamma$. However, the quantity $\nu_{E}=\nu(1-\gamma^{2})=0.027$
turns out to be the same for both $j=1$ and $2$. This gives the
estimated value of $3.4$ ps for the rotational energy relaxation
time $\tau_{E}$. This value correlates with the experimentally measured
long time anisotropy decay times of $2\div3$ ps \cite{tok04,tok05,bak00,bak05,fay05,wie04}.

Note, finally, that a close interrelation between the hydrogen bond
dynamics and rotation dynamics has repeatedly been emphasized in the
literature. The NKSM values of the {}``persistence time'' of the
oscillatory angular momentum CF (Fig. 1a) and the rotational energy
relaxation time, $\sim150$fs and $3.4$ps, correlate quite well with
the estimations for the continuous and intermitted hydrogen bond lifetimes
\cite{ski03,luz00}. This observation is consistent with the physical
picture of molecular rotation underlying the NKSM approach. The in-cage
librations, which can cause bond breakings, manifest themselves on
the time scale of the {}``persistence time'' of the angular momentum
CF. On the other hand, the bond rupturings are accompanied by subsequent
bond reformings, since the molecule does not leave its local cage.
A ``true'' cleavage of the bond can occur on the $\tau_{E}$ timescale
since, within the NKSM description, the rotational energy relaxation
occurs due to a hopping to a new position of the local equilibrium
or restructuring the local potential well.

\section{Conclusion}
We have developed a NKSM description of the orientational relaxation
in hydrogen-bonding liquids. Within the NKSM, molecular rotation is
governed by two relaxation parameters $\nu$ and $\gamma$ (which
describe the intensity and mechanism of dissipation), as well as by
the memory function $g(t)$ (\ref{g(t)}), which is responsible for
the short-time dynamics. Alternatively, the relaxation parameters
$\nu$ and $\gamma$ are uniquely determined through the angular momentum
and energy relaxation times $\tau_{J}$ (\ref{TauJ}) and $\tau_{E}$
(\ref{TauE}). Once a set of the parameters is selected, the NKSM
allows to calculate any rotational CF or OCF of interest. Keeping
in mind a considerable success of the KSM in reproducing molecular
reorientation in gases \cite{BurTe,gel00}, the results of the present
work demonstrate that the NKSM can be used for the description
and interpretation of the orientational relaxation in a condensed
phase, from rarefied gases with binary collisions, through dense fluids
to hydrogen-bonding liquids. 

The NKSM suggests the short-time relaxation of the OCFs in the hydrogen-bonding
liquids is described by an algebraic $\sim t^{-3/2}$ contribution. This algebraic behavior is
 modified  by more rapid and non-monotonous dynamics, which is induced by the memory effects.
 The long-time decay
of the OCFs is exponential, although non-diffusional. It is governed
by the rotational energy relaxation time, $\tau_{E}$.
Our results are contrary to standard belief that
the angular momentum CF determines molecular reorientation
in the hindered rotation limit, and the first cumulant expression,
Eq. (\ref{GtShort1}), delivers the leading contribution into the
OCF. Our results indicate, that knowing $C_{J}(t)$ is not enough
to predict OCFs for hydrogen-bonding liquids, since the long-time
behavior of OCFs is governed by the rotational energy CF, $C_{E}(t)$.

It is a conventional practice to fit various  experimental or simulated CFs
via a linear combination of several real (if the CF decays monotonously)
or complex (if the CF exhibits oscillatory behavior) exponents. According
to the present analyses, the OCF in hydrogen-bonding liquids contains an algebraic $\sim t^{-3/2}$ contribution. 
This finding suggests that the following  fitting formulas  for the angular momentum CF, \begin{equation}
C_{J}(t)=a_{1}\exp\{-\nu_{1}t\}+\exp\{-\nu_{2}t\}(a_{2}\cos(\Omega t)+a_{3}\sin(\Omega t)),\label{FitCJ}\end{equation}
 and for the OCF,\[
G^{j}(t)=b_{1}\exp\{-\nu_{1}t\}+\exp\{-\nu_{2}t\}(b_{2}\cos(\Omega t)+b_{3}\sin(\Omega t))\]
\begin{equation}
+b_{4}\exp\{-\nu_{3}t\}\left[1+b_{5}t\right]^{-3/2}+b_{6}\exp\{-\nu_{4}t\},\label{Fit}\end{equation}
can be more physically motivated ($a_{i},\, b_{i}$, $\nu_{i}$ and
$\Omega$ being certain real-valued parameters). Indeed, Eq. (\ref{FitCJ})
is nothing else than a formal solution of Eq. (\ref{CJ}) in the case
of two-exponential memory function (\ref{g(t)}). As to the OCF (\ref{Fit}),
the first three terms with coefficients $b_{1},\, b_{2},\, b_{3}$
describe the angular momentum induced short-time rapid decay and oscillations.
The term which is proportional to $b_{4}$ is responsible for the algebraic
contribution, and the last term governs the long-time exponential
decay. 

\begin{acknowledgments}
The authors are grateful to Joseph Loparo for sending them the numerical
data on the simulated OCFs and measured anisotropies, which have
been published in \cite{tok05}. M. F. G. thanks Alexander Blokhin
for numerous stimulating and useful discussions. 
\end{acknowledgments}

\section{Appendix A. Recursive relations for linear and spherical top OCFs}

After being Laplace transformed, Eq. (\ref{OCF1}) reads:\begin{equation}
-\rho_{B}(\mathbf{J})+s\tilde{G}^{j}(\mathbf{J},s)=-i(\Lambda_{\mathbf{\Omega}}+\hat{\Lambda}_{\mathbf{J}})\tilde{G}^{j}(\mathbf{J},t)-\nu \tilde{g}(s)\{\tilde{G}^{j}(\mathbf{J},s)-\int d\mathbf{J}'T(\mathbf{J}|\mathbf{J}')\tilde{G}^{j}(\mathbf{J}',s)\}.\label{OCFL}\end{equation}
Since Eq. (\ref{OCFL}) depends on $s$ parametrically, the method
of its solution in the Markovian limit ($\tilde{g}(s)=1$), which has been
been developed in \cite{sack,gel96,gel96a,gel00}, is directly applicable
to the present case also. One has merely consider the complex quantity
$\nu g(s)$ as the generalized relaxation rate. We therefore present
the final expressions for the calculation of the first and second
rank OCFs. 

Their Laplace images of the spherical top OCFs can be calculated via
the formula\begin{equation}
\tilde{G}^{j}(s)=(1+2b_{0})/s.\label{Gb0}\end{equation}
For $j=1$, the coefficient $b_{0}$ can be retrieved from the simple
three-term recursive formula \cite{sack,gel96}\begin{equation}
\frac{1}{s}\delta_{m0}=\frac{4m+10}{\sigma_{m+1}}b_{m+1}-\left\{ \frac{2m+3}{\sigma_{m+1}}+\frac{2m+2}{\sigma_{m}}+\zeta_{m}\right\} b_{m}+\frac{m}{\sigma_{m}}b_{m-1},\label{Lj1}\end{equation}
$\delta_{m0}$ being the Kronecker delta. The value of $b_{0}$ for
the second rank OCF can be extracted from the system of coupled recursive
relations for the coefficients $b_{m}$ and $d_{m}$ \cite{gel96}:\begin{equation}
\frac{1}{s}\delta_{m0}=-\left\{ \frac{6}{\sigma_{m}}+\zeta_{m}\right\} b_{m}+\frac{12}{\sigma_{m+1}}b_{m+1}\label{Lj2a}\end{equation}
\[
-\frac{4m}{\sigma_{m}}d_{m-1}+\left\{ \frac{8m-14}{\sigma_{m}}+\frac{8m-31}{\sigma_{m+1}}\right\} d_{m}+\frac{-16m^{2}+74m+12}{(m+1)\sigma_{m+1}}d_{m+1};\]

\textcompwordmark{}

\textcompwordmark{}\begin{equation}
0=\frac{m}{\sigma_{m}}b_{m-1}-\left\{ \frac{2m-1}{\sigma_{m}}+\frac{2m-1}{\sigma_{m+1}}\right\} b_{m}+\frac{4m}{\sigma_{m+1}}b_{m+1}\label{Lj2b}\end{equation}
\[
+\frac{5m}{\sigma_{m}}d_{m-1}-\left\{ \frac{14}{\sigma_{m}}+\frac{4+10m}{\sigma_{m+1}}+\zeta_{m}\right\} d_{m}+\frac{4m(9+5m)}{(m+1)\sigma_{m+1}}d_{m+1}.\]
Here \begin{equation}
\sigma_{m}\equiv s+\nu \tilde{g}(s)(1-\gamma^{2m}),\,\,\,\zeta_{m}\equiv s+\nu \tilde{g}(s)(1-\gamma^{2m+1}).\label{sisi}\end{equation}
The solution of the recursive relations for the first rank OCF (\ref{Lj1})
can be expressed in the continued fraction form \cite{sack}, while
the solution of Eqs. (\ref{Lj2a}) and (\ref{Lj2b}) for the second-rank
OCF can be given in terms of the matrix $2\times2$ continued fractions
\cite{ris}.

The first and second rank OCFs for linear rotors can be evaluated
very similarly, through the simple three-term recursive formulas.
Namely, the Laplace images of the OCFs can also be computed via Eq.
(\ref{Gb0}). For $j=1$, the coefficient $b_{0}$ must be determined
by the formula \cite{sack,gel00} \[
\frac{1}{s}\delta_{m0}=\frac{4m+8}{\sigma_{m+1}}b_{m+1}-\left\{ \frac{2m+2}{\sigma_{m+1}}+\frac{2m+2}{\sigma_{m}}+\zeta_{m}\right\} b_{m}+\frac{m}{\sigma_{m}}b_{m-1},\]
while the second-rank OCF ($j=2$) can be computed through \cite{gel00}\[
\frac{3}{s}\delta_{m0}=\frac{16m+32}{\sigma_{m+1}}b_{m+1}-\left\{ \frac{8m+10}{\sigma_{m+1}}+\frac{8m+6}{\sigma_{m}}+\zeta_{m}\right\} b_{m}+\frac{4m}{\sigma_{m}}b_{m-1}.\]

\section{Appendix B. Orientational relaxation in case of perfect angular momentum
reorientation}

If we neglect the memory effects ($g(t)\rightarrow\delta(t)$) and
put $\gamma=-1$, then Eq. (\ref{OCF1}) can be solved analytically: 

\begin{equation}
G^{j}(t)=\frac{1}{2j+1}\int_{0}^{\infty}dJ\rho_{B}(J)\label{Gt1}\end{equation}
\[
\times\left(1+\exp\{-\nu t\}\sum_{k=1}^{j}\left[(1+\nu/\omega_{k})\exp\{\omega_{k}t\}+(1-\nu/\omega_{k})\exp\{-\omega t\}\right]\right).\]
 The frequencies are explicitly defined as follows\begin{equation}
\omega_{k}=\sqrt{\nu^{2}-k^{2}J^{2}}.\label{Wk}\end{equation}
Several important properties of Eq. (\ref{Gt1}) are to be discussed.
OCF (\ref{Gt1}) possesses a stationary asymptote: $G^{j}(t\rightarrow\infty)=(2j+1)^{-1}$,
which is identical to the free OCF asymptote. This can be easily understood:
reversion of the angular momentum is equivalent to the reversion of
the sense of molecular rotation. Therefore a sequence of forward-backward
rotations is equivalent to a single free rotation. This asymptote
is solely caused by dynamic effects (in-cage librations, compare with
\cite{eva95}) rather than by external potentials (see, e.g., \cite{sza84,ham05,fay05}).
The librational motion itself is caused, of course, by in-cage potentials
but they do not enter explicitly into our analysis. If $\nu\rightarrow0$,
Eq. (\ref{Gt1}) reproduces the free spherical top OCF. If we take
the opposite limit $\nu\rightarrow\infty$, then $G^{j}(t)\rightarrow1$
since a large number of small-angle forward-backward rotations causes
no net reorientation. In the hindered rotation limit ($\tau_{J}\ll1$)
Eq. (\ref{Gt1}) reduces to (\ref{Gt2}).

Using the explicit form of the free linear rotor OCF \cite{ste69}
one can easily derive the linear rotor counterpart of Eq. (\ref{Gt2}):\begin{equation}
G^{j}(t)=\left(d_{00}^{j}(\frac{\pi}{2})\right)^{2}+2\sum_{k=1}^{j}\left(d_{0k}^{j}(\frac{\pi}{2})\right)^{2}\left[1+\frac{k^{2}}{\nu_{l}}t\right]^{-1},\label{Gtlin}\end{equation}
 $d_{km}^{j}(\beta)$ being the reduced Wigner function \cite{var89}.
It is well known that the exponent $d$ of the long-time hydrodynamic
tails $t^{-d/2}$ of the angular velocity CFs of Brownian particles
is determined by the dimensionality of the rotation space: $d=3$
for any spherical, symmetric or asymmetric top while $d=2$ for a
linear rotor (\cite{fel97,mas97} and references therein). This general
statement holds true in the present case also, and OCF (\ref{Gtlin})
possesses a $\sim t^{-1}$ tail. This means that the algebraic contribution
to OCF (\ref{Gtlin}) decays slower than its $t^{-3/2}$ counterpart
in the spherical top OCF (\ref{Gt2}). Furthermore, $d_{00}^{j}(\pi/2)=0$
for odd $j$. Thus, the odd-ranked OCFs do not possess the stationary
contribution and decay faster then the even-ranked OCFs.

\clearpage

\begin{figure}
\includegraphics[keepaspectratio,totalheight=10cm,angle=270]{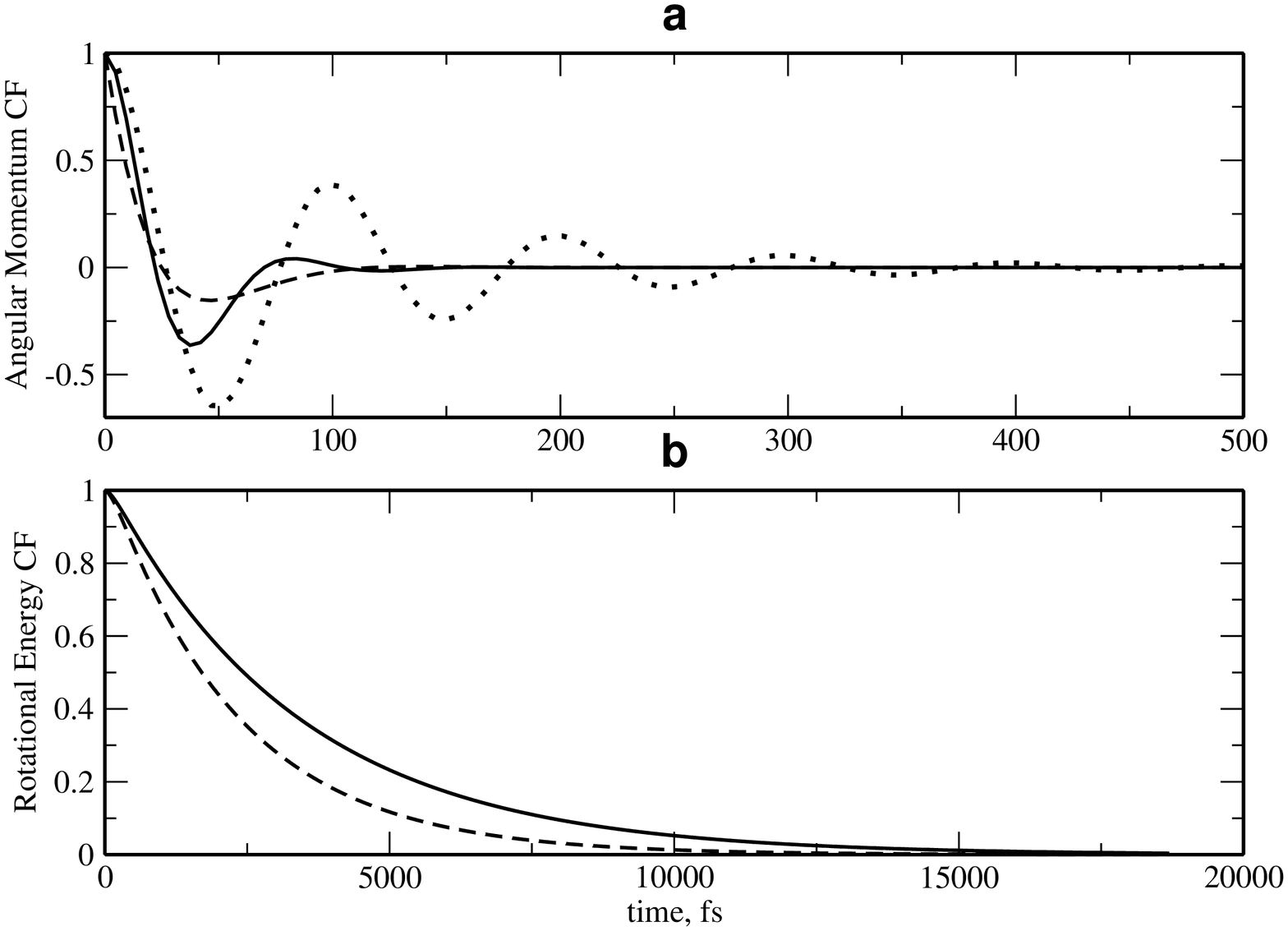}
\caption{Angular momentum (a) and energy (b) CFs for $\nu=16$, $\gamma=-0.99915$
(that is $\nu_{J}=32$ and $\nu_{E}=0.027$), $\sigma=0.2$ and $\lambda_{1}=0.7$.
The dashed, full and dotted lines correspond to $\lambda_{2}=1000$,
$10$ and $3$, respectively. On the scale of the figure, the rotational
energy CFs for $\lambda_{2}=$ $10$ and $3$ are indistinguishable.
}
\end{figure}

\clearpage

\begin{figure}
\includegraphics[keepaspectratio,totalheight=12cm,angle=270]{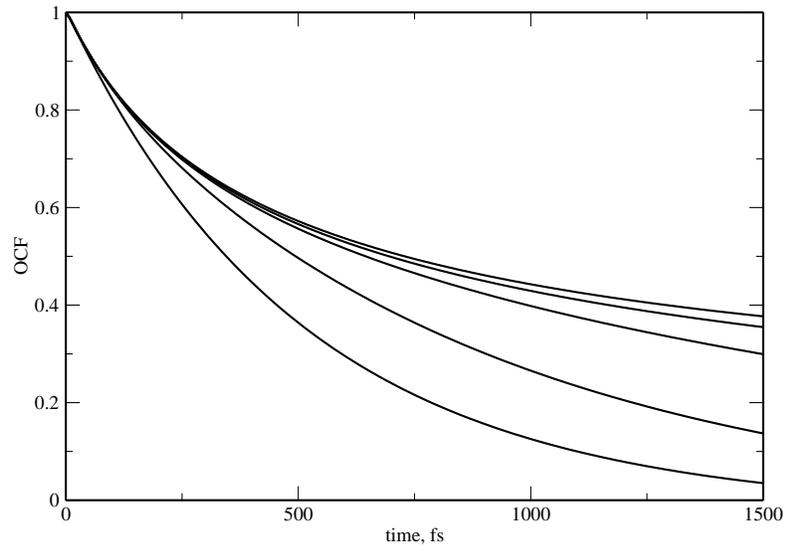}
\caption{ The second rank OCFs in the Markovian limit for the relaxation
rate rate $\nu_{J}=32$. From bottom to top, the curves correspond
to $\gamma=1$, $-0.99$, $-0.999$, $-0.9999$ and $1$. 
} 
\end{figure}

\clearpage

\begin{figure}
\includegraphics[keepaspectratio,totalheight=12cm,angle=270]{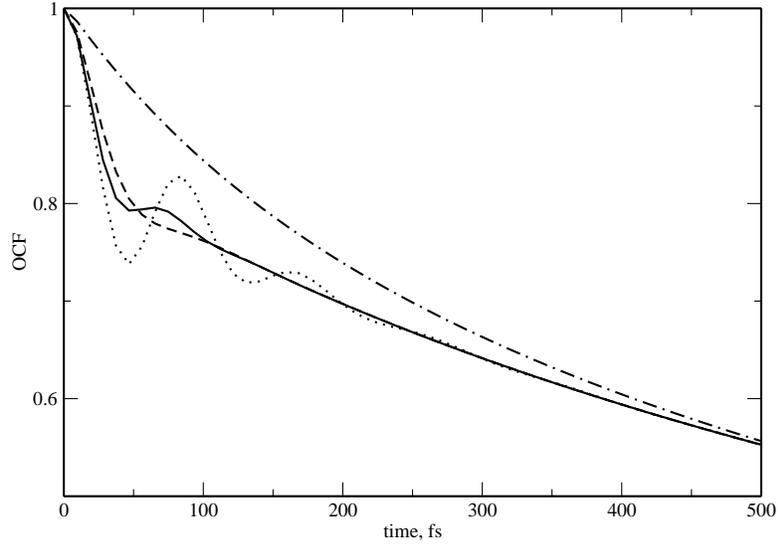}
\caption{
 The influence of the memory effects on the second rank OCFs.
$\nu_{J}=32$, $\gamma=$ $-0.999$, $\sigma=0.2$ and $\lambda_{1}=0.7$;
$\lambda_{2}=1000$ (dashed lines), $10$ (full lines) $3$ (dotted
lines). The dash-dotted curve depicts the OCF in the Markovian limit.}
\end{figure}

\clearpage
\begin{figure}
\includegraphics[keepaspectratio,totalheight=12cm,angle=270]{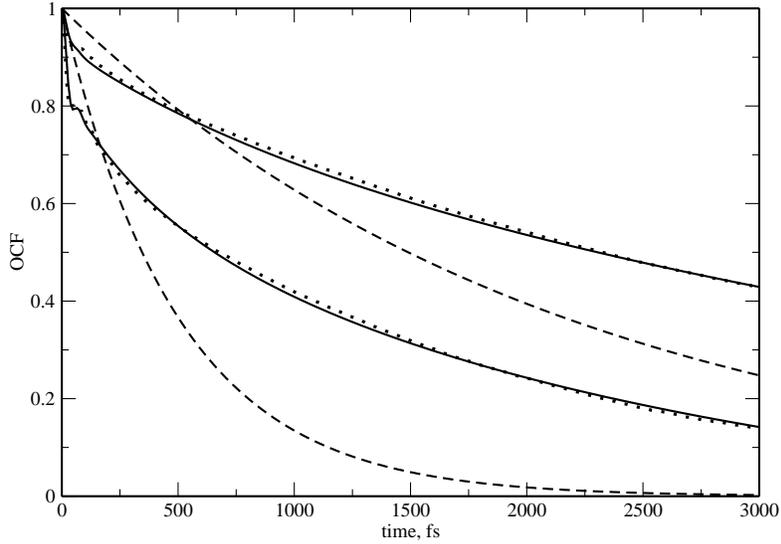}
\caption{
 Comparison of the simulated OCFs with those calculated within
the NKSM. The dotted lines correspond to the first (upper
curve) and second (lower curve) rank OCFs which were simulated for
$HOD$ in liquid $D_{2}O$ at a room temperature \cite{tok05}. The black
curves are computed for $\nu_{J}=46$, $\gamma=$ $-0.999415$, $\sigma=0.2$,
$\lambda_{1}=0.4$, $\lambda_{2}=10$ ($j=1$); $\nu_{J}=32$, $\gamma=$
$-0.99915$, $\sigma=0.2$, $\lambda_{1}=0.7$, $\lambda_{2}=10$
($j=2$). The OCFs calculated via the diffusion equation (\ref{dif})
are depicted by dashed lines.  }
\end{figure}

\clearpage
\begin{figure}
\includegraphics[keepaspectratio,totalheight=12cm,angle=270]{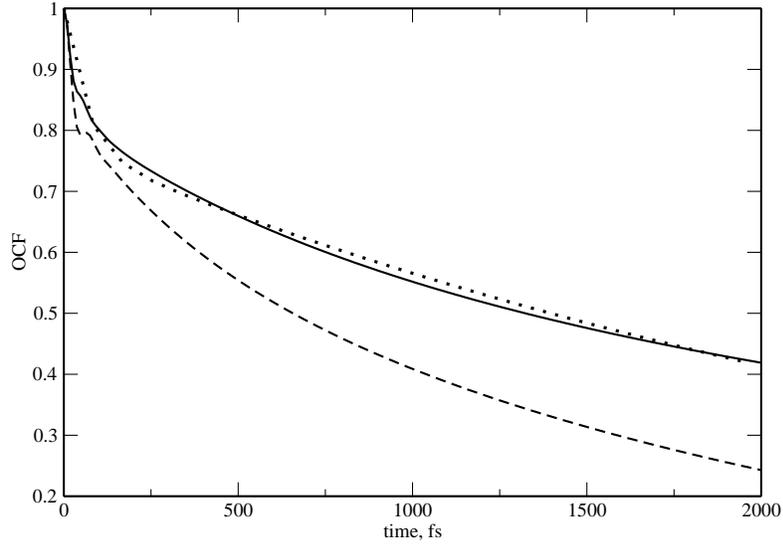}
\caption{ Comparison of the experimental OCFs with those calculated
within the NKSM. The dotted line reproduces the best fit
to the experimental anisotropy decay \cite{tok05}. The solid black
curve shows the second-rank OCF computed within the NKSM for $\nu_{J}=63$,
$\gamma=$ $-0.99995$, $\sigma=0.2$, $\lambda_{1}=0.3$, $\lambda_{2}=10$.
The dashed line reproduces the second rank NKSM OCFs from Fig. 4. }
\end{figure}

\clearpage
\begin{figure}
\includegraphics[keepaspectratio,totalheight=10cm,angle=270]{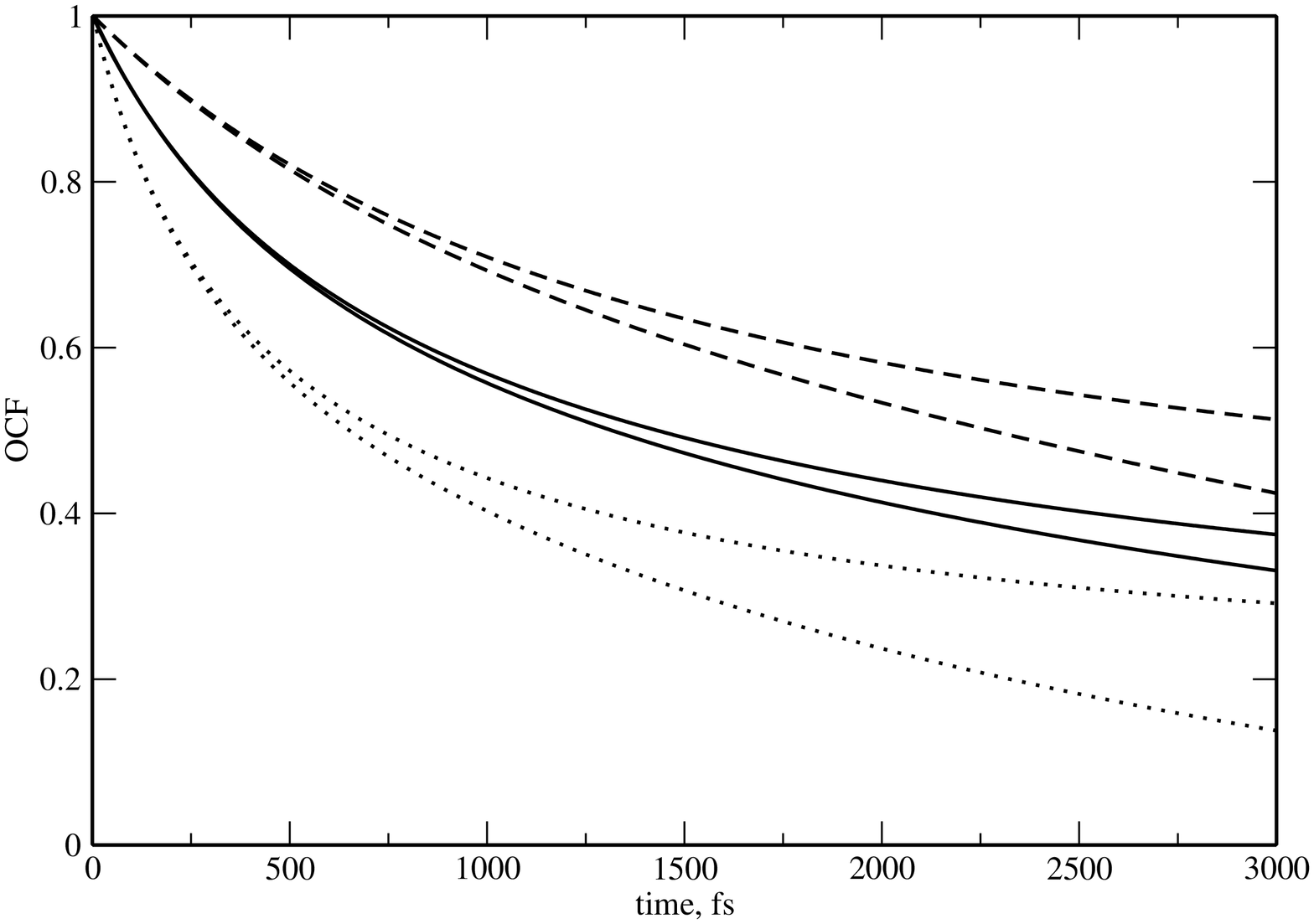}
\caption{
Elucidation of the algebraic contributions into OCFs. For
each couple of the curves, the upper one is calculated via Eq. (\ref{Gt2})
and the lower one is computed within the Markovian limit of the NKSM.
The dotted ($j=2$, $\nu_{J}=32$, $\gamma=$ $-0.99915$) and dashed
($j=1$, $\nu_{J}=46$, $\gamma=$ $-0.999415$) curves correspond
to the best-fit NKSM OCFs from Fig. 4, and the solid curves ($j=2$,
$\nu_{J}=63$, $\gamma=$ $-0.99995$) correspond to the best-fit
NKSM OCFs from Fig. 5.}
\end{figure}


\begin{thebibliography}{20}
\bibitem{foot2}Strictly speaking, these hydrodynamic long-time tails are associated
with local structures and confining potentials in liquids, but these
effects are evidently not taken into account within the NKSM description.
\end{thebibliography}
\end{document}